\documentstyle[preprint,aps]{revtex}
\begin{document}
\draft
\title{Transport of magnetoexcitons in single and coupled
quantum wells}
\author{Yu.~E.~Lozovik$^a$ \thanks{E-mail: lozovik@isan.troitsk.ru },
A.~M.~Ruvinsky$^b$}
\address{$^a$Institute of Spectroscopy, Academy of Sciences of Russia,
142092,Troitsk, Moscow region, Russia}
\address{$^b$Department of Theoretical Physics,
Moscow State Institute for Steel and Alloys,
Leninsky prosp.~4, 117936, Moscow, Russia}

\date{\today}
\tighten
\narrowtext
\maketitle
\begin{abstract}

The transport relaxation time $\tau (P)$
and the mean free path of magnetoexcitons
in single and coupled quantum wells
are calculated
(~$P$ is the magnetic momentum of the
magnetoexciton~). We present the results
for magnetoexciton scattering in a
 random field due to (~i~)  quantum well width fluctuations,
(~ii~) composite fluctuations
and (~iii~) ionized impurities.
The time $\tau(P)$ depends nonmonotonously
on $P$
in the case (~ii~) and
in the cases (~i~), (~iii~)
for $D/l$ smaller than some critical
value (~$D$ is the interwell separation,
$l=\sqrt{\hbar c/eH}$ is the magnetic length~).
For $D/l\gg 1$ the transport relaxation time 
increases
monotonously with $P$.

The magnetoexciton mean free
path $\lambda (P)$ has a maximum at $P\ne 0$
in the cases (~i~), (~iii~).
It decreases
with increasing $D/l$.
The mean free path calculated for the case (~ii~)
may have two maxima. One of them
disappears with the variation of the
random fields parameters.
The maximum of $\lambda (P)$ increases
with $H$ for
types (i,iii) of scattering processes and decreases in the case
(ii).

\end{abstract}


\newpage
\section{Introduction}
Electron(e)-hole(h) systems
in
coupled quantum wells
particularly in strong magnetic fields, have
attracted  considerable interest over
the past few years~[1-7],
long after the
prediction of the superfluidity in the system
manifesting itself in a sense as a superconductivity
in the "electron" and "hole" quantum wells or wires~\cite{a8}
(~see also~[9,10] and references therein; the magnetic field
effects were analyzed in~[11-15]~).
An interesting drag effects in this system
was also studied~[16,17].
The application of strong
transverse magnetic fields
increases the effective exciton mass and
exciton binding energy~\cite{lerner,ruvin} and thus
 effect on  the
conditions of the superfluid
phase formation~\cite{berman}.

Indirect excitons or pairs of
spatially separated e and h may also condense in a variety of
different phases. Some of them are analogous
to the phases in the system of bulk excitons
~\cite{chen,keld}.
The existence of these phases needs
sufficiently long exciton life time 
in comparison with the thermalization time.
Experimental investigations~\cite{butler,a6} of  the
photoluminescense of spatially-separated electron-hole systems
in transverse magnetic fields in coupled quantum wells~(CQW)
discovered dramatical
reductions of the lifetimes of indirect magnetoexcitons
 with increasing magnetic field strength for
$B>7\mbox{T}$ and at low temperatures, $T< 1\mbox{K}$.
Indirect excitons are characterized by large radiative
lifetimes because the
spatial confinement of electrons and holes in different
quantum wells (~QW~) suppresses the  e-h annihilation rate
due to the small overlap of electron and hole wave functions.
One can conclude therefore that the total exciton
lifetime is
determined by the
relaxation time of exciton
transport
to the recombination centers which strongly
depends on the magnetic field. Exciton-exciton
collisions and indirect-direct exciton conversion
induced by the collisions
are not essential
in the dilute exciton gas.

Two-dimensional (~2D~) exciton diffusion and localization
in random fields
at $B=0$ were considered in [20-23].
The transport properties
of the bulk excitons were studied in [24-26].
Low temperature transport of 2D excitons
in {\it double} QWs in the limiting
cases of zero magnetic field and magnetic quantum limit
is considered in~\cite{dzub}.
In the present paper we study
the effects of the interface roughness
(~IFR~), impurities in CQWs and
local concentration fluctuations in
$A^{III}B^{V}$ solid solution
on 2D direct and indirect
magnetoexciton transport
at low temperatures.

We show that the transport
momentum relaxation time $\tau$ of indirect magnetoexcitons
essentially depends on the magnetoexciton
momentum ${\bf P}$ and interwell separation $D$
between electron and hole QWs.
It occurs that
the transport momentum relaxation times of
direct and indirect magnetoexciton scattering
on IFR
are nonmonotoneous functions of ${\bf P}$ and
have maximum $\tau_{max}$ at nonzero magnetic
momentum and small interwell separation
due to the nonmonotonous dependence of  the
magnetoexciton group velocity
and interaction matrix element
on ${\bf P}$.
The calculated transport
relaxation time behavior is in qualitative
agreement with the experimental
data~\cite{butler}
in the limiting case of high magnetic fields.

The organization of this paper is as
follows. The transport momentum
relaxation time and mean free path of magnetoexciton
scattering on IFR in QW and
in CQWs are calculated
respectively in Sec.~II and Sec.~IV.
In Sec.~III we consider magnetoexciton
transport in the random field of the
local
concentration fluctuations in
$A^{III}B^{V}$ solid solution.
The transport properties
of magnetoexcitons in the presence of
randomly distributed ionized impurities are
considered in Sec.~V. Conclusions are presented
in Sec.~VI.

\section{Direct magnetoexciton IFR scattering}
The scattering potential for excitons can
arise due to the local
fluctuations in the QW width and have the form\
~\cite{ray,dzub}
\begin{equation}
\label{sqw}
V({\bf r}_e,{\bf r}_h)=V_e({\bf r}_e)+V_h({\bf r}_h),
\end{equation}
$$
V_{e,h}({\bf r})=\alpha_{e,h} (\xi_1({\bf r})-\xi_2({\bf r})),
$$
where
$\alpha_{e,h}=\frac{\partial E_{e,h}^{(0)}}{\partial d}$,
$d$ is the QW width,
$E_{e(h)}^{(0)}$ is the energy of the  e(h) lowest
level in the conduction (~valence~)
band, $\xi_{1,2}({\bf r})$ are local QW width
fluctuations on different interfaces. The
fluctuations
$\xi_{1}({\bf r})$ and
$\xi_{2}({\bf r})$ are statistically
independent, while the fluctuations on the
same interface are described by Gaussian autocorrelation
functions, i.e.
\begin{equation}
\label{xi}
\ll\xi_i (r_1)\xi_j (r_2)\gg=\delta_{ij}\Delta_{i}^2\exp\left(
-\frac{(r_1-r_2)^2}{2\Lambda_i^2}
\right),
\end{equation}
where $\Delta_i$ is the amplitude of fluctuations
$\xi_i$, $\Lambda_i$ is the in-plane correlation
length.
Further we consider the case of high transverse
magnetic fields (~$l\ll a_{e,h}$; in this case
magnetoexciton states can be classified
by Landau level quantum numbers $n$ and $m$
~\cite{lerner,ruvin}~),
thin QWs (~$d^2\ll a_{e,h}l$; in this case
the energy of e and h size quantization is
much greater than the exciton energies~) and
small width fluctuations $\Delta_i\ll d$
(~$a_{e,h}=\epsilon \hbar^2/m_{e,h}e^2$ are the
effective Bohr radii of electron (~e~) and hole (~h~),
$\epsilon =(\epsilon_1+\epsilon_2)/2$,
$\epsilon_{1,2}$
are the dielectric susceptibilities of
the media surrounding e and h QWs~).
The value $\Delta_i$ can have the order
of one monolayer height $\approx 3\AA$
(~see~[28-30]~). The correlation length
$\Lambda_i$ essentially depends
on the QW growth method~\cite{tata}.
One can consider two characteristic
exciton transport regimes in dependence
on the value $\Lambda_i/r_{exc}$
(~$r_{exc}$ is the exciton size in the QW plane layer which
is proportional
to $l$ in high magnetic fields~\cite{lerner,ruvin}~).
For $\Lambda_i\approx r_{exc}$ excitons
may be localized in minima of local potential fluctuations
with average size
$\Lambda_i$. Hence localized exciton states are
formed~\cite{butov2}, analogous to states in
quantum dots~\cite{dots}
(~excitons in 'natural' quantum dots~).
In this case the low-
temperature exciton transport is apparently thermally-activated
between states in different minima of random potential~\cite{taka}.
In sections II and IV of the paper we consider the opposite
case $\Lambda_i\gg r_{exc}$ which takes place in
structures $AlGaAs-GaAs$
with smooth interfaces.
(~see~[28-30]~). More strictly the criterium
of the interface smoothness in the framework
of the exciton problem is
\begin{equation}
\label{buda}
r_{exc}\sqrt{\ll\nabla V^2\gg}\ll E_I,
\end{equation}
where $E_I$ is the energy of exciton ionization.
For $D\ll l$ we get from inequality~(\ref{buda})
\begin{equation}
\Lambda\gg\frac{a_{e,h}^zl^2\Delta}{d^3}\sqrt{2\pi^3}
\end{equation}
and for $D\gg l$
\begin{equation}
\label{poi}
\Lambda\gg\frac{a_{e,h}^zlD\Delta}{d^3}\sqrt{2\pi^3}
\end{equation}
(~$a_{e,h}^z=\epsilon \hbar^2/m_{e,h}^ze^2$ are the
effective Bohr radii of e and h in the z-direction~).

The transport momentum relaxation time
of magnetoexcitons is calculated in the Born
approximation, which is valid in the case of
$\hbar^2/M_{exc}\Lambda^2\gg
\Delta\hbar^2/\pi^2m^{z}_{e,h}d^3$, i.e.
\begin{equation}
\label{born}
\pi\sqrt{\frac{d^3}{\Delta}\frac{m^{z}_{e,h}}{M_{exc}}}\gg
\Lambda
\end{equation}
With $D$ and $H$ increasing the effective
magnetoexciton mass
$M_{exc}$ increases~\cite{lerner,ruvin}.
Hence the upper limit
of the  correlation length variation decreases.
The magnetoexciton energy $E_I$ decreases (increases) with $D$ (~$H$~).
The lower limit of $\Lambda$ variation
increases (~decreases~) with $D (H)$.

The magnetoexciton relaxation time in Born
approximation is given by
~\cite{ray,agran,dzub}
\begin{equation}
\label{time}
\tau^{-1}(P)=\frac{2\pi}{\hbar}\sum\limits_{P'}
\ll |<P'|V({\bf r}_e,{\bf r}_h)|P>|^2\gg
(1-cos(\phi_{PP'}))\delta ({\cal E}(P)-{\cal
E}(P')),
 \end{equation}
where ${\cal E}(P)$ is the
dispersion law of magnetoexcitons in QW (~direct
magnetoexciton~) in the lowest Landau level
($n=m=0$)~\cite{lerner}:
\begin{equation}
\label{mono}
{\cal E}(P)=\frac{1}{2}\hbar\omega_c-\frac{e^2}{\epsilon l}
\sqrt{\frac{\pi}{2}}I_0\left(\frac{P^2l^2}{4\hbar^2}\right)
\exp\left(-\frac{P^2l^2}{4\hbar^2}\right)
\end{equation}
where $\omega_c=eH/\mu c$ is the cyclotron energy,
$\mu =m_em_h/(m_e+m_h)$ is the in-plane reduced mass,
${\bf P}$ is the magnetoexciton momentum
(~which is a conserved quantity for an isolated
magnetoexciton in an ideal system~),
$I_n(x)$ are modified Bessel functions.
The scattering on IFR between two magnetoexciton
states
$(n=m=0,{\bf P})$ and  $(n=m=0,{\bf P}')$
is described by the matrix element
~(\ref{sqw})
$$
<{\bf P}'\mid V({\bf r}_e,{\bf r}_h)\mid {\bf P}>=\frac{1}{S}
\exp\left(-({\bf P}'-{\bf P})^2l^2/4\hbar^2\right)
$$
\begin{equation}
\left(V_e({\bf P}'-{\bf P})\exp\left(il^2{\bf H}[{\bf P},{\bf
P}']/2\hbar^2H\right)+V_h({\bf P}'-{\bf P})\exp\left(-il^2{\bf
H}[{\bf P},{\bf P}']/2\hbar^2H\right)\right),
\end{equation}
then
$$
\ll\mid<{\bf P}'\mid V({\bf r}_e,{\bf r}_h)\mid {\bf P}>\mid^2\gg=
$$
\begin{equation}
\label{avs}
\frac{\pi}{2S}\left(\alpha_e^2+\alpha_h^2+
2\alpha_e\alpha_h
\cos\left(l^2[{\bf P},{\bf P}']/\hbar^2\right)\right)
\sum\limits_{i=1,2}
\exp\left(-({\bf P}'-{\bf P})^2l^2\beta_i/2\hbar^2\right)
\left(\Delta_i\Lambda_i\right)^2,
\end{equation}
where $\beta_i=1+\Lambda_i^2/l^2$.

Using Eqs.~(\ref{avs}) and~(\ref{time})
the following expression for the direct
magnetoexciton relaxation time is obtained:
\begin{equation}
\label{tsqw}
\tau^{-1}(P)=\tau_e^{-1}(P)+\tau_h^{-1}(P)+\tau_{eh}^{-1}(P)\; ,
\end{equation}
where
$$
\tau_{e,h}^{-1}(P)=\frac{\pi}{4\hbar^3}\left|\frac{\partial {\cal E}(P)}
{\partial P^2}\right|^{-1}\alpha_{e,h}^2\sum\limits_{i=1,2}
(\Delta_i\Lambda_i)^2\exp(-P^2l^2\beta_i/\hbar^2)
$$
\begin{equation}
\left(
I_0(P^2l^2\beta_i/\hbar^2)-I_1(P^2l^2\beta_i/\hbar^2)
\right)\;
\end{equation}
and
$$
\tau_{eh}^{-1}(P)=\frac{\pi\alpha_e\alpha_h}{2\hbar^3}
\left|\frac{\partial
{\cal E}(P)} {\partial P^2} \right|^{-1}
\sum\limits_{i=1,2}
(\Delta_i\Lambda_i)^2\exp(-P^2l^2\beta_i/\hbar^2)
$$
\begin{equation}
\label{liok}
\left(
I_0(P^2l^2\sqrt{\beta_i^2-1}/\hbar^2)-
\frac{\beta_i}{\sqrt{\beta_i^2-1}}
I_1(P^2l^2\sqrt{\beta_i^2-1}/\hbar^2)
\right)
\end{equation}
For $l\ll a_{e,h}$ and $d^2\ll a_{e,h}l$
transverse motion of e and h along the magnetic
field does not lead to qualitative changes
of the exciton spectrum ~\cite{lerner,ruvin},
so analogously to~\cite{ray} we assume that
$\alpha_{e,h}=-\pi^2\hbar^2/m^{z}_{e,h}d^3$.

We have calculated the transport momentum relaxation time
(~Fig.~1~)
of direct excitons at $H=20\;\mbox{Oe}$ in $GaAs/Al_xGa_{1-x}As$ QWs
using the following values of parameters:
$d=40\AA$, $\Lambda_1=100\AA$, $\Lambda_2=30\AA$,
$\Delta_1=\Delta_2=3\AA$,
e and h effective masses in H-direction
are
$m^{z}_{e}=1.1(1.3)$ and $m^{z}_{h}=0.75(0.34)$
in $AlAs(GaAs)$ (~see Ref.~\cite{salma}~).
The function $\tau (P)$
is {\it nonmonotonous} and has a maximum at
$Pl=2.6\hbar$. It occurs because the
scattering probability (~$1/\tau$~)
(~due to the density of states in the final state -
see Eq.~(\ref{tsqw},\ref{liok})~)
is inversely proportional to the
magnetoexciton group velocity
\begin{equation}
\frac{\partial {\cal E}(P)}{\partial P^2}=
\frac{V_g(P)}{2P}
\end{equation}
The function $V_g(P)$ has a {\it maximum}
which {\it corresponds to the saddle point} of ${\cal E}(P)$.
With $H$ increasing $\tau_{max}$ increases
and shifts to larger magnetic momenta.
Expression~(\ref{tsqw}) has the following
asymptotics:
\begin{equation}
\tau (P)=\frac{\hbar e^2l}{\sqrt{2\pi}\epsilon}
\frac{1}{(\alpha_e+\alpha_h)^2}
\cases{
\left(
(\Delta_1\Lambda_1)^2+
(\Delta_2\Lambda_2)^2\right)^{-1}
,&$P=0$,\cr
8\left(
\frac{(\Delta_1\Lambda_1)^2}{\beta_1^{3/2}}+
\frac{(\Delta_2\Lambda_2)^2}{\beta_2^{3/2}}
\right)^{-1}
,&$Pl/\hbar\gg 1$,\cr
}
\end{equation}

The magnetoexciton mean free path is treated as
$\lambda (P)=v(P)\tau (P)$ (~Fig.~2~),
where $v(P)=\frac{\partial {\cal E}(P)}{\partial P}$.
For $P=0$ we obtain $\lambda (0)=0$ ($v(0)=0$) and
for $Pl/\hbar\gg 1$ we get
\begin{equation}
\lambda =\frac{4\sqrt{2}\hbar^2 e^4}
{\sqrt{\pi}\epsilon^2P^2l}
\frac{1}{(\alpha_e+\alpha_h)^2}
\left(
\frac{(\Delta_1\Lambda_1)^2}{\beta_1^{3/2}}+
\frac{(\Delta_2\Lambda_2)^2}{\beta_2^{3/2}}
\right)^{-1}
\end{equation}
For $Pl/\hbar\ll 1$ and $Pl/\hbar\gg 1$
the magnetoexciton mean free path is typically smaller
than the correlation length of disorder
$\Lambda$. This
situation means magnetoexciton localization
~\cite{future}. Magnetic field increase
leads to  increasing $\lambda_{max}$.

\section{Composition fluctuations in QW}

Random potential
in QW can arise due to local concentration fluctuations
of the components of QW structure~\cite{gevor,ram}.
The interaction of excitons with composition
fluctuations has the form~\cite{bar}
\begin{equation}
\label{ol}
V({\bf
r}_e,{\bf r}_h)=\alpha_e \xi({\bf r}_e)- \alpha_h
\xi({\bf r}_h) , \end{equation}
where
$\alpha_{e,h}=\frac{1}{N}\frac{\partial E_{e,h}^{(0)}}{\partial x}$,
$x$ is the mean relative concentration of A-atoms,
$N$ is the concentration of lattice sites where two
types of atoms may arrange,
$\xi ({\bf r})$ is the surplus concentration of one of 
the components
of solid solution characterized by the
correlation function~\cite{bar}
 \begin{equation}
\label{}
\ll \xi ({\bf r}_1)\xi ({\bf r}_2)\gg=
Nx(1-x)\delta ({\bf r}_1-{\bf r}_2)
\end{equation}
Further we estimate the criterion of
applicability of the
Born approximation
using the method given in~\cite{bar}.
The mean number of type-$A$ atoms in the
volume $R^2d$ (~d is the QW width~)
is $xNR^2d$. The typical surplus number is
of the order of
$(xNR^2d)^{1/2}$.
Hence the type-$A$ atoms share variation
is
 \begin{equation} \delta
x=\frac{(xNR^2d)^{1/2}}{xNR^2d} \end{equation}

Thereby the exciton potential well is
\begin{equation}
\label{yama}
V=\alpha\frac{(xNR^2d)^{1/2}}{xNR^2d}
\end{equation}
(~$\alpha =\alpha_e-\alpha_h$~) arises in the
volume $R^2d$. For $\alpha_e=\alpha_h$ the 'exciton well'
is equal to zero. This means that in the last case
the interaction~(\ref{ol}) does not produce
a change in the total exciton energy.
An analogous situation arises also for direct
exciton scattering by ionized impurities (~Sec.~V~).
The Born approximation is valid under the condition
$V\ll \hbar^2/M_{exc}R^2$, i.e.

\begin{equation}
\frac{\hbar^3}{RM_{exc}}\gg\alpha\sqrt{\frac{x}{Nd}}
\end{equation}

The matrix element~(\ref{ol}) for transitions
between  states $(n=m=0,{\bf P})$ and\\$(n=m=0,{\bf P}')$
has the form
$$
<{\bf P}'\mid V({\bf r}_e,{\bf r}_h)\mid {\bf P}>=\frac{1}{S}
\exp\left(-({\bf P}'-{\bf P})^2l^2/4\hbar^2\right)
\xi({\bf P}'-{\bf P})
$$
\begin{equation}
\left(\alpha_e\exp\left(il^2{\bf H}[{\bf P},{\bf
P}']/2\hbar^2H\right)-\alpha_h\exp\left(-il^2{\bf
H}[{\bf P},{\bf P}']/2\hbar^2H\right)\right)
\end{equation}
Then
$$
\ll\mid<{\bf P}'\mid V({\bf r}_e,{\bf r}_h)
\mid {\bf P}>\mid^2\gg=
$$
\begin{equation}
\label{avs1}
\frac{g}{S}\left(\alpha_e^2+\alpha_h^2-
2\alpha_e\alpha_h
\cos\left(l^2[{\bf P},{\bf P}']/\hbar^2\right)\right)
\exp\left(-({\bf P}'-{\bf P})^2l^2/2\hbar^2\right),
\end{equation}
where $g=Nx(1-x)$.

Using Eqs.~(\ref{avs1}) and (\ref{time}) we obtain the
transport relaxation time
$$
\tau^{-1}_{cf}(P)=
\frac{g}{\hbar^3}\left|\frac{\partial {\cal E}(P)}
{\partial P^2}\right|^{-1}
\exp(-P^2l^2/\hbar^2)
$$
\begin{equation}
\label{tsqw1}
\left(\frac{\alpha_{e}^2+\alpha_h^2}{2}
\left(
I_0(P^2l^2/\hbar^2)-I_1(P^2l^2/\hbar^2)
\right)
+\alpha_e\alpha_h\left(\frac{
P^2l^2}{\hbar^2}-1\right)\right)
\end{equation}
The function $\tau_{cf}$ is {\it nonmonotonous} (~Fig.~3 a,b~).
Such behavior is due to {\it interplay between the maximum}
of $\partial {\cal E}/\partial P^2$  and the {\it minimum}
of the rest part of~(\ref{tsqw1}).
With $\alpha_h/\alpha_e$ increasing the minimum
of $\tau_{cf}(P)$ appears at $P\ne 0$.
Further increase of the parameter
$\alpha_h/\alpha_e$ leads to the minimum
disappearing.

The asymptotics of expression~(\ref{tsqw1}) are
\begin{equation}
\tau_{cf}(P)=\frac{\hbar e^2l\sqrt{\pi}}{2^{3/2}g\epsilon}
\cases{
(\alpha_e-\alpha_h)^{-2}
,&$P=0$,\cr
8\pi(\alpha_e^2+\alpha_h^2)^{-1}
,&$Pl/\hbar\gg 1$\cr
}
\end{equation}

Taking into account values of $\alpha_e-
\alpha_h$ (~see~\cite{bar} and references therein~) we
obtain $\tau_{cf}\sim 10^{-13}-10^{-11} \mbox{c}$.
Therefore one can conclude
that in sufficiently narrow QWs the magnetoexciton
transport relaxation time defiened by  exciton
scattering on
composition fluctuations have the same
order as the relaxation time for scattering on IFR.

Using Eqs.~(\ref{mono}) and~(\ref{tsqw1}) we
obtain the magnetoexciton mean free path
as function of $P$ (~Fig.~4 a,b~).
For sufficiently small values of the
parameter $\alpha_e/\alpha_h$
it has a maximum at $P\ne 0$
analogously to the case of magnetoexciton
IFR scattering. With $\alpha_e/\alpha_h$
increasing the second maximum appears
and then disappears. The value
$\lambda_{max}$ decreases with 
increasing $H$. 
It indicates that in strong magnetic fields
excitons may be localized in random fields of composite
fluctuations.

\section{Indirect magnetoexciton IFR scattering.}
The scattering potential of a magnetoexciton with
spatially separated e and h has the form
\begin{equation}
V({\bf r}_e,{\bf r}_h)=\alpha_e (\xi_1({\bf r}_e)
-\xi_2({\bf r}_e))+
\alpha_h (\xi_3({\bf r}_h)-\xi_4({\bf r}_h)) ,
\end{equation}
where $\xi_{1}$, $\xi_2$ (~$\xi_3$, $\xi_4$~)
 are e(h) QW width
fluctuations on the lower and upper interfaces,
$\alpha_{e,h}=\frac{\partial E_{e,h}^{(0)}}{\partial d_{e,h}}$,
$d_{e,h}$ are e and h QW widths.
As before we assume that
$\xi_i$ in different interfaces are statistically independent and have
Gaussian autocorrelation functions ~(\ref{xi}) on the same interface.
This
is possible under the condition that the interwell separation $D$
between e and h
QWs is
larger than the amplitudes $\Delta$ of fluctuations on the nearest
interfaces of e and h QWs.  The opposite case $D\le \Delta$
can be realized
in {\it double} QWs and for $D=0$ it is
considered in~\cite{dzub}.

For the case of interest we have
$$
\ll\mid<{\bf P}'\mid V(r_e,r_h)\mid {\bf P}>\mid^2\gg=\frac{\pi}{2S}
\left\{\alpha_e^2
\sum\limits_{i=1,2}
\exp\left(-({\bf P}'-{\bf P})^2l^2\beta_i/2\hbar^2\right)
\left(\Delta_i\Lambda_i\right)^2+
\right.
$$
\begin{equation}
\label{qqq}
\left.
\alpha_h^2
\sum\limits_{i=3,4}
\exp\left(-({\bf P}'-{\bf P})^2l^2\beta_i/2\hbar^2\right)
\left(\Delta_i\Lambda_i\right)^2\right\}
\end{equation}
The relaxation time of indirect magnetoexcitons
is obtained from Eqs.~(\ref{qqq}) and
~(\ref{time}) in the form
\begin{equation}
\tau^{-1}(P)=\tau_e^{-1}(P)+\tau_h^{-1}(P)\; ,
\end{equation}
where
$$
\tau_{e,h}^{-1}(P)=\frac{\pi}{4\hbar^3}
\left|\frac{\partial {\cal E}(P,D)}
{\partial P^2}
\right|
^{-1}\alpha_{e,h}^2\sum\limits_{i=1,2(3,4)}
(\Delta_i\Lambda_i)^2\exp(-P^2l^2\beta_i/\hbar^2)
$$
\begin{equation}
\left(
I_0(P^2l^2\beta_i/\hbar^2)-I_1(P^2l^2\beta_i/\hbar^2)
\right),
\end{equation}
${\cal E}(P,D)$ is the dispersion law
of an indirect magnetoexciton in the state
with quantum numbers $n=m=0$~\cite{ruvin}
\begin{eqnarray}
\label{v}
{\cal E}({\cal P},{\cal D})=\frac{1}{2}\hbar\omega_c
-\frac{e^2}{\epsilon l}\frac{1}{\sqrt{2}}
f({\cal D},{\cal P})
\end{eqnarray}
\begin{eqnarray}
f({\cal P},{\cal D})=
\sum\limits_{k=0}^{\infty}P_{2k}(g)
\frac{(-1)^{k}(2k)!}{2^{2k}(k!)^{2}}\biggl\{
\left(\frac{2}{{\cal D}^{2}+{\cal P}^{2}}\right)^{k+1/2}
\gamma\left(k+1,
\frac{{\cal D}^{2}+{\cal P}^{2}}{2}\right)+\nonumber
\end{eqnarray}
\begin{equation}
\left(\frac{{\cal D}^{2}+{\cal P}^{2}}{2}\right)^{k}
\Gamma\left(-k+\frac{1}{2},
\frac{{\cal D}^{2}+{\cal P}^{2}}{2}\right)\biggr\},
\end{equation}
where ${\cal D}=D/l$, ${\cal P}=Pl/\hbar$
$g=\left(1+\left(\frac{{\cal P}}{{\cal D}}\right)^2\right)^{-1/2}$,
$\Gamma(a,x)$ is the complete gamma-function,
$\gamma(a,x)$
is the incomplete gamma-function,
$P_k(x)$ is the Legendre polynominal~\cite{prudnik}.
The transport relaxation time of an indirect
magnetoexciton with an electron in AlAs QW and
a hole in GaAs QW is presented in Fig.~1.
The parameters of e and h QWs
coincide with the parameters used in Sec.~2 for
transport relaxation time
calculation
of direct magnetoexciton.
The dependence of $\tau (P,D)$ on $P$
is nonmonotonous.
With increasing  $D/l$ the value of maximum
$\tau_{max}$ decreases
due to decreasing indirect magnetoexciton
group velocity. For
$D/l$ larger than some critical value the relaxation
time $\tau (P,D)$ became a monotonously
increasing function of $P$ (~Fig.~1~).
For sufficiently large $D$ we expect
that a minimum of $\tau (P,D)$ appears.
But for large interwell
separation our results may not be valid
due to~(\ref{poi},\ref{born}).

For a small magnetic momentum $Pl/\hbar\ll 1$ the
relaxation time is
\begin{equation}
\label{tyty}
\tau(0,D)=
\frac{2\hbar^3}{\pi}\frac{1}{M({\cal D})}
\left(\alpha_e^2\sum\limits_{i=1,2}
\left(\Delta_i\Lambda_i\right)^2
+
\alpha_h^2\sum\limits_{i=3,4}
\left(\Delta_i\Lambda_i\right)^2\right)^{-1},
\end{equation}
where
\begin{equation}
\label{massa}
M({\cal D})=\frac{M_o}{(1+{\cal D}^2)e^{\frac{{\cal D}^2}{2}}
erfc\left(\frac{{\cal D}}{\sqrt{2}}\right)
-{\cal D}\sqrt{\frac{2}{\pi}}}
\end{equation}
is the indirect magnetoexciton effective mass
in the state with quantum numbers $n=m=0$ (~see~\cite{ruvin}~),
$M_o=\frac{2^{3/2}\epsilon\hbar^2}{e^2l\sqrt{\pi}}$
is the direct magnetoexciton effective mass
in the lowest Landau level~\cite{lerner}.

With increasing
magnetic field the relaxation time
~(\ref{tyty}) decreases as
$ 1/\sqrt{H}$ at $D\ll l$
and as $ 1/H^2$ at $D\gg l$ in agreement
with experimentally observed data~\cite{butler}.
For $Pl/\hbar\gg 1$ and $PD/\hbar\gg 1$ the magnetoexciton
dispersion law~(\ref{v}) is
${\cal E}\approx 0.5\hbar\omega_c-e^2\hbar/\epsilon Pl^2$. Hence for
large magnetic momentum we obtain
\begin{equation}
\label{bigp}
\tau \approx
\frac{4\sqrt{2}\hbar e^2l}{\epsilon\sqrt{\pi}}
\left(\alpha_e^2\sum\limits_{i=1,2}
\frac{\left(\Delta_i\Lambda_i\right)^2}{\beta^{3/2}_i}
+\alpha_h^2\sum\limits_{i=3,4}
\frac{
\left(\Delta_i\Lambda_i\right)^2}{\beta^{3/2}_i}\right)^{-1}
\end{equation}
Expression~(\ref{bigp})
decreases as $ 1/\sqrt{H}$
and is independent of P and D.

It is interesting to note that $\tau (P,D)$ and
$\lambda (P,D)$ may have several extrema
for magnetoexcitons in the excited states
due to nonmonotonous behavior of
${\cal E}_{nm}(P,D)$~\cite{lerner,ruvin}.
Magnetoexciton
dispersion laws calculated in~\cite{lerner,ruvin}
have $2n+2$ saddle points in the states with quantum
numbers $(m\ne 0,n)$ and $2n+1$ saddle points
in the case $(m=0,n)$. {\it Each saddle point corresponds
to the maximum of the function $|V_g|$}.
With $D/l$ increasing the maximum of $|V_g|$
decreases
and dissapears at some critical $D/l$~\cite{ruvin}.
This variation
may lead to the existences of {\it
several maxima} of $\tau (P)$
{\it which disappear with} 
{\it increasing} $D$
(~nonmonotonous behavior of interaction
matrix elements must also be
take into account~).

The mean free path of indirect magnetoexciton
is shown in Fig.~2. The value of $\lambda_{max}$
decreases with increasing $D/l$.
If $P=0$ we get $\lambda (0,D)=0$ because $v(0,D)=0$.
If
$Pl/\hbar\gg 1$
the mean free path is $\lambda\sim\frac{1}{P^2}$.
With $H$ increasing $\lambda_{max}$ increases.
\section{Ionized impurities in CQW}

Here we consider the situation when ionized
Coulomb impurities $q_1$ and $q_2$
are randomly distributed in both  QWs
with the corresponding concentrations $n_1$ and $n_2$.
The interaction of excitons with impurities
in CQW is described by the potential energy:
\begin{equation}
\label{cobel}
V({\bf r}_e,{\bf r}_h)=V_e({\bf r}_e)+V_h({\bf r}_h)\; ,
\end{equation}
where
\begin{equation}
V_e({\bf r}_e)=\sum\limits_i\frac{-eq_1}{\epsilon|{\bf r}_e-{\bf r}_i|}+
\sum\limits_j\frac{-eq_2}{\epsilon
\sqrt{({\bf r}_e-{\bf r}_j)^2+D^2}}
\end{equation}
\begin{equation}
V_h({\bf r}_h)=
\sum\limits_i\frac{eq_1}{\epsilon\sqrt{({\bf r}_h-{\bf r}_i)^2+D^2}}+
\sum\limits_j\frac{eq_2}{\epsilon|{\bf r}_h-{\bf r}_j|}
\end{equation}
The impurities $q_1$ and $q_2$ occupy
places with coordinates ${\bf r}_i$ and ${\bf r}_j$ in CQW.
For $D=0$ all impurities are located in one QW.
The averaged squared matrix element of
exciton-impurity interaction~(\ref{cobel}) has
the following
form
$$
\ll\mid <{\bf P}'|V({\bf r}_e,{\bf r}_h)|{\bf P}>\mid^2\gg=
\exp\left(-\frac{({\bf P}'-{\bf P})^2l^2}{2\hbar^2}
\right)
\left(\frac{2\pi\hbar e}{{\bf P}'-{\bf P}}\right)^2
\frac{(n_1q_1^2+n_2q_2^2)}{S\epsilon^2}
$$
\begin{equation}
\label{ipppi}
\left\{1+\exp\left(-\frac{2D|{\bf P}'-{\bf
P}|}{\hbar}\right)-2\exp\left(-\frac{D|{\bf P}'-{\bf P}|}{\hbar}
\right)
\cos\left(\frac{[{\bf P},{\bf P}']l^2}{\hbar^2}\right)\right\}
\end{equation}

Using Eqs.~(\ref{time}), (\ref{ipppi}), we get the inverse
relaxation time in the form (~Fig.~5~)
$$
\frac{1}{\tau_i(P,D)}=
\frac{\pi}{2\hbar}\left(\frac{e}{\epsilon P}\right)^2
\left|\frac{\partial {\cal E}(P,D)}{\partial P^2}\right|^{-1}\exp
\left(-\frac{P^2l^2}{\hbar^2}\right)
(n_1q_1^2+n_2q_2^2)
\int\limits_0^{2\pi}\exp\left(\frac{P^2l^2\cos\phi}{\hbar^2}\right)
$$
\begin{equation}
\label{iii}
\left(1+\exp\left(-\frac{2PD\sqrt{2-2\cos\phi}}{\hbar}\right
)-2\exp\left(-
\frac{PD\sqrt{2-2\cos\phi}}{\hbar}\right)
\cos\left(\frac{P^2l^2}{\hbar^2}\sin\phi\right)
\right)d\phi \end{equation}

For $P=0$ the transport relaxation time is
\begin{equation}
\label{qaqa}
\tau_i(0,D)=\frac{\hbar^3}{4\pi^2}\frac{\epsilon^2}
{e^2D^2M({\cal D})}\frac{1}{(n_1q_1^2+n_2q_2^2)}
\end{equation}
The divergence in Eq.~(\ref{qaqa}) at small $P$
and $D$ arises due to the use
of unscreened Coulomb potential.
The relaxation time
$\tau_i(0,D)$ decreases with increasing $H$ because of
increasing
magnetoexciton effective mass~(\ref{massa}).
For $Pl/\hbar\gg 1$ and $PD/\hbar\gg 1$ the relaxation
time~(\ref{iii}) in asymptotic approximation does not depend on $P,D$:
\begin{equation}
\tau_i =\frac{\hbar\epsilon}{\sqrt{2}\pi^{3/2}l(n_1q_1^2+
n_2q_2^2)}
\end{equation}
and increases as $\sqrt{H}$.
The mean free path of a magnetoexciton in the presence
of impurities in CQW is shown in Fig.~6.
The peak of $\lambda_i(P,D)$ decreases with 
increasing $D/l$. The calculated value of
the mean free path
$\lambda_i(P,D)$ controlled by ionized
impurities
is greater for the same $P$ and $D$ than $\lambda (P,D)$
controlled by interface roughness
scattering
(~calculated in Sec.~II and Sec.~IV~).

\section{Conclusions}
In the present paper low temperature transport
properties of direct and indirect magnetoexcitons
in CQW are considered. The transport relaxation time
and mean free path of magnetoexcitons are calculated
in Born approximation for various scattering
processes. In the case $Pl/\hbar\ll 1$
the transport of direct magnetoexcitons
is found to be limited by the
interface roughness scattering
or composition fluctuations because of $\tau_i\to\infty$.
The transport relaxation time of direct
magnetoexciton scattering on composite
fluctuations depends nonmonotonously on $P$.
The transport relaxation times $\tau $, $\tau_i$
are nonmonotonous functions of $P$ for parameter
$D/l$ smaller than some critical value.
The relaxation time
$\tau (P,D)$ has a maximum at $Pl\approx 2.6/\hbar$
which decreases with increasing $D$
and disappears
at $D$
greater than some critical value.
The maximal time
$\tau_{max} $ increases
with increasing magnetic field  strength and shifts to larger
magnetic momenta.
For large $D/l$ the quantities
$\tau$ and $\tau_i$ increase monotonously with $P$.
The mean free path of direct and indirect magnetoexcitons
is a nonmonotonous function of $P$.
It has  a maximum at $P\ne 0$
which decreases with $D$. For $H$
increasing $\lambda_{max}$
increases in the case of IFR scattering and decreases
due to scattering on local composition fluctuations.

{\bf Acknowledgements.}\\
Authors are obliged to A.I. Belousov
for useful discussions of results.

The work was supported by the Russian Foundation of
Basic Reseach, INTAS and the Program
"Fundamental  Spectroscopy".

\newpage
\begin{large}
\begin{center}
Captions to Figures for Article\\
"Transport of magnetoexcitons in single and coupled
quantum wells"\\
by Yu.~….~Lozovik and €.~Œ.~Ruvinsky
\end{center}

Fig.~1.~
Transport relaxation time
of direct
 (~dots line~) and indirect (~solid line~)
magnetoexciton
scattering
on interface terraces
in coupled quantum wells at
$H=2\cdot 10^5\mbox{Oe}$
as the function of magnetic momentum $P$.
Lines 1,2,3,4,5 correspond to
interwell sepration
$D/l=0.1;0.5;1;2;3$.
$P$ is given in units $\hbar/l$.

Fig.~2.~
Mean free path of
direct (~broken line~) and indirect (~solid
lines~)
magnetoexciton
scattering
on interface terraces
in coupled quantum wells at
$H=2\cdot 10^5\mbox{Oe}$
as a function of magnetic momentum $P$.
Lines 1,2,3 correspond to $D/l=0.1;0.5;1$.

Fig.~3.~
Transport relaxation time
of direct magnetoexciton
scattering
on composition fluctuations
at
$H=2\cdot 10^5\mbox{Oe}$
as a function of magnetic momentum $P$.\\
a) Lines 1,2,3,4,5 correspond to
$\alpha_h/\alpha_e=0;0.25;0.5;0.75;1$;
$\tau_{cf}$ is given in units
$e^2\hbar l/g\alpha_e^2\epsilon$,
$P$ is given in units $\hbar/l$.\\
b) Lines 1,2,3,4 correspond to
$\alpha_h/\alpha_e=1.75;2;4;5$.

Fig.~4.\\
Mean free path of
direct
magnetoexciton
scattering
on composition fluctuations
at
$H=2\cdot 10^5\mbox{Oe}$
as a function of magnetic momentum $P$.\\
a) Lines 1,2,3,4,5,6 correspond to
$\alpha_h/\alpha_e=0;0.25;0.5;0.75;1.25;1.5$;
$\lambda_{cf}$ is given in units
$e^4l/g\alpha_e^2\epsilon^2$.\\
b) Lines 1,2,3,4,5 correspond to
$\alpha_h/\alpha_e=1.75;2;4;5;6$.

Fig.~5.~
The inverse transport relaxation time of
direct (~broken line~) and indirect (~solid
lines~)
magnetoexciton scattering
on ionized impurities
in coupled quantum wells at
$H=2\cdot 10^5\mbox{Oe}$
as a function of magnetic momentum $P$.
Lines 1,2,3,4 correspond to $D/l=0.1;0.3;0.5;1$;
impurity concentrations
$c_1=c_2= 10^{10}
\mbox{cm}^{-2}$.

Fig.~6.~
Mean free path of
direct (~broken line~) and indirect (~solid
lines~)
magnetoexciton
scattering
on ionized impurities
in coupled quantum wells at
$H=2\cdot 10^5\mbox{Oe}$
as a function of magnetic momentum $P$.
Lines 1,2,3 correspond to $D/l=0.1;0.5;1$;
impurity concentrations
$c_1=c_2=10^{10}
\mbox{cm}^{-2}$.

\end{large}

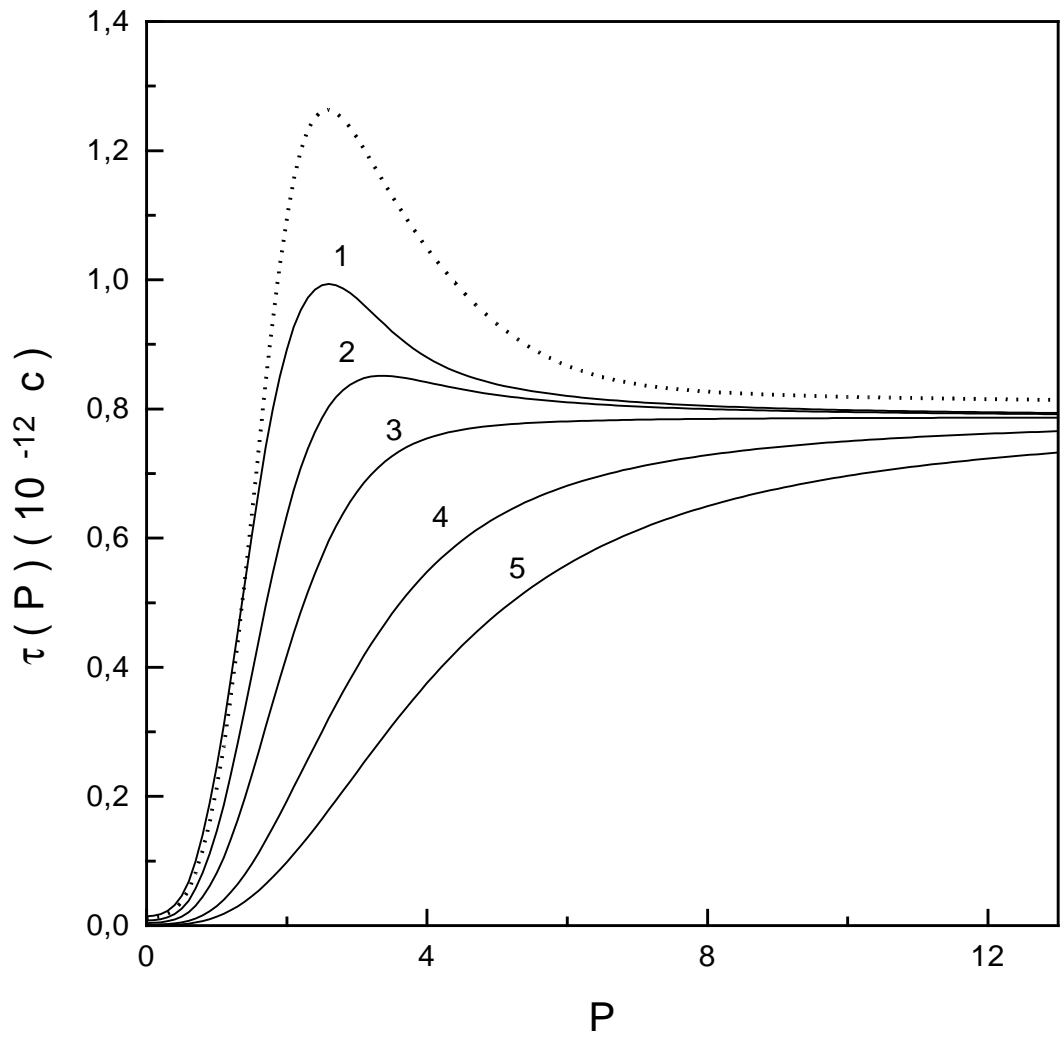
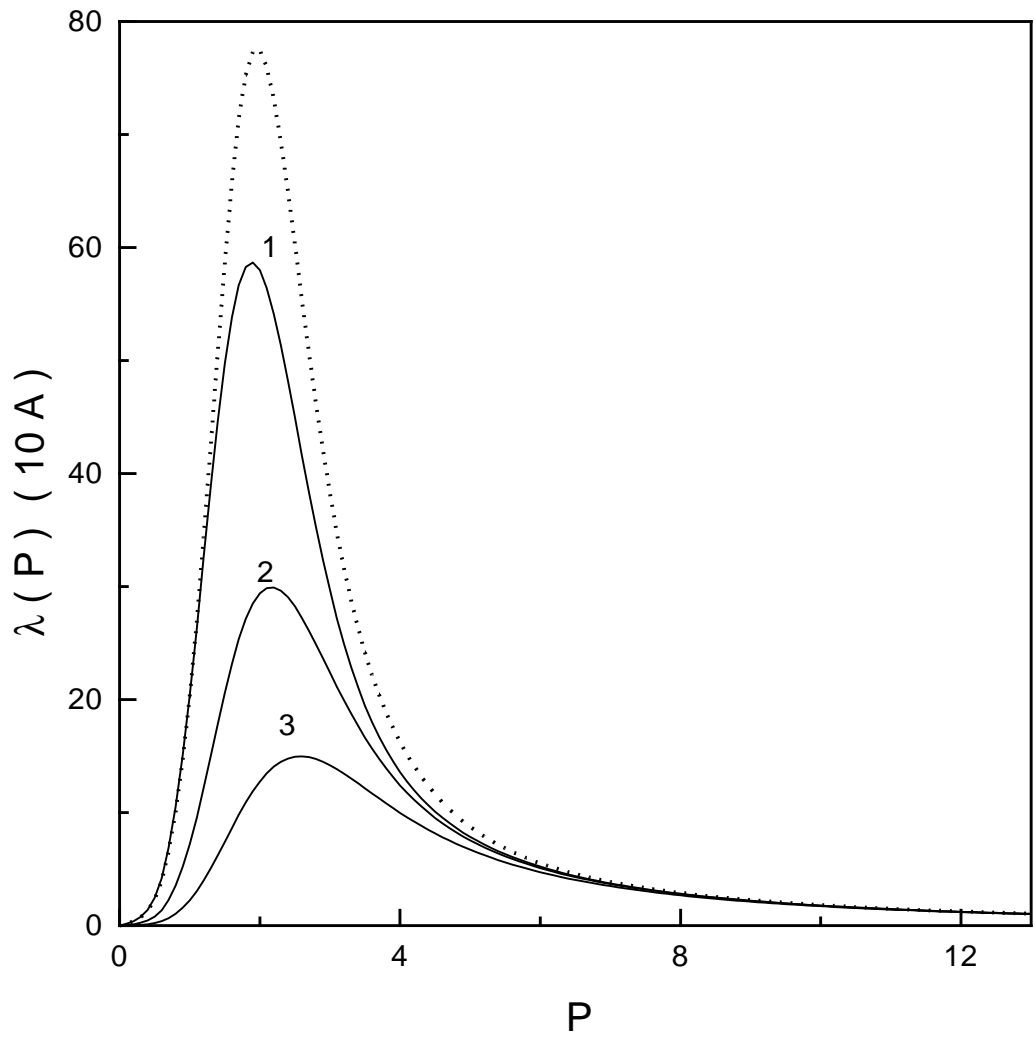
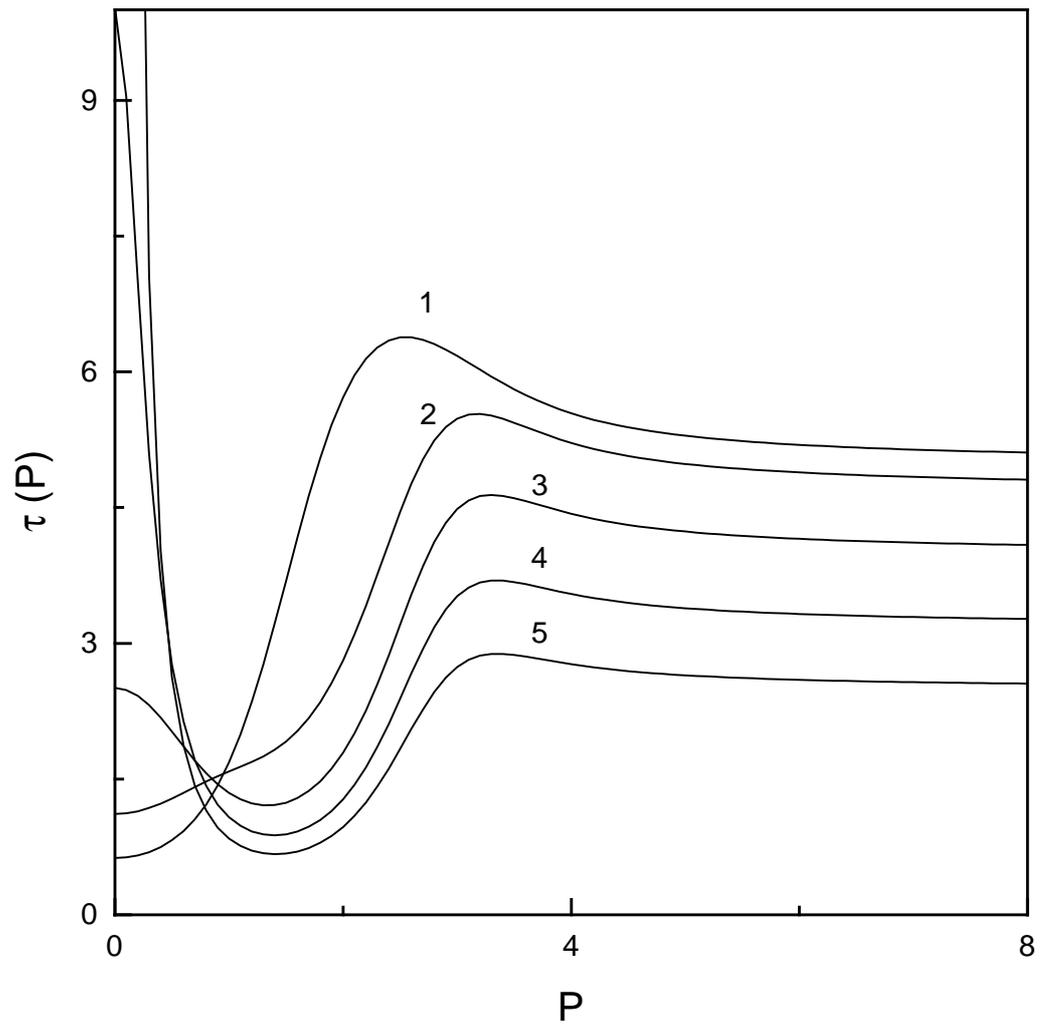
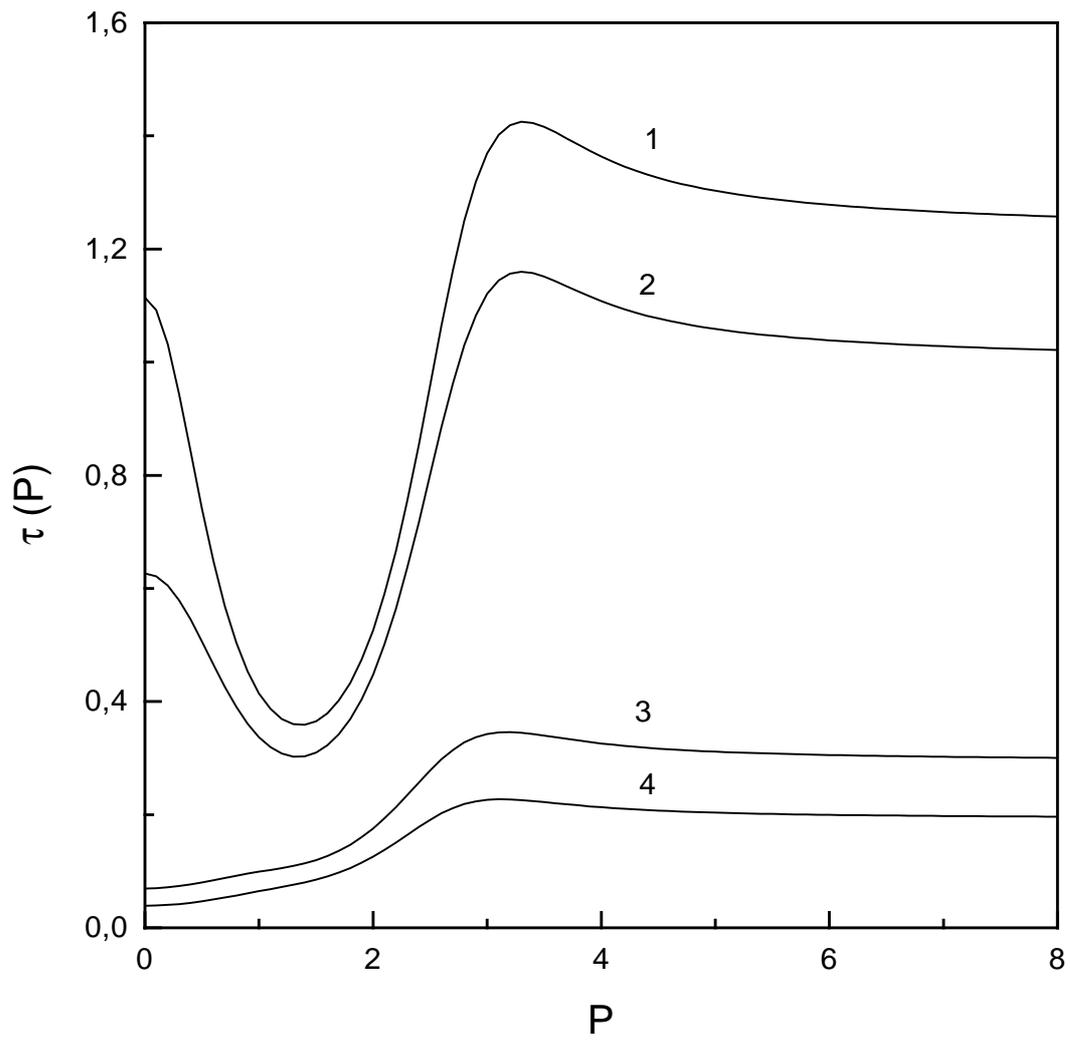
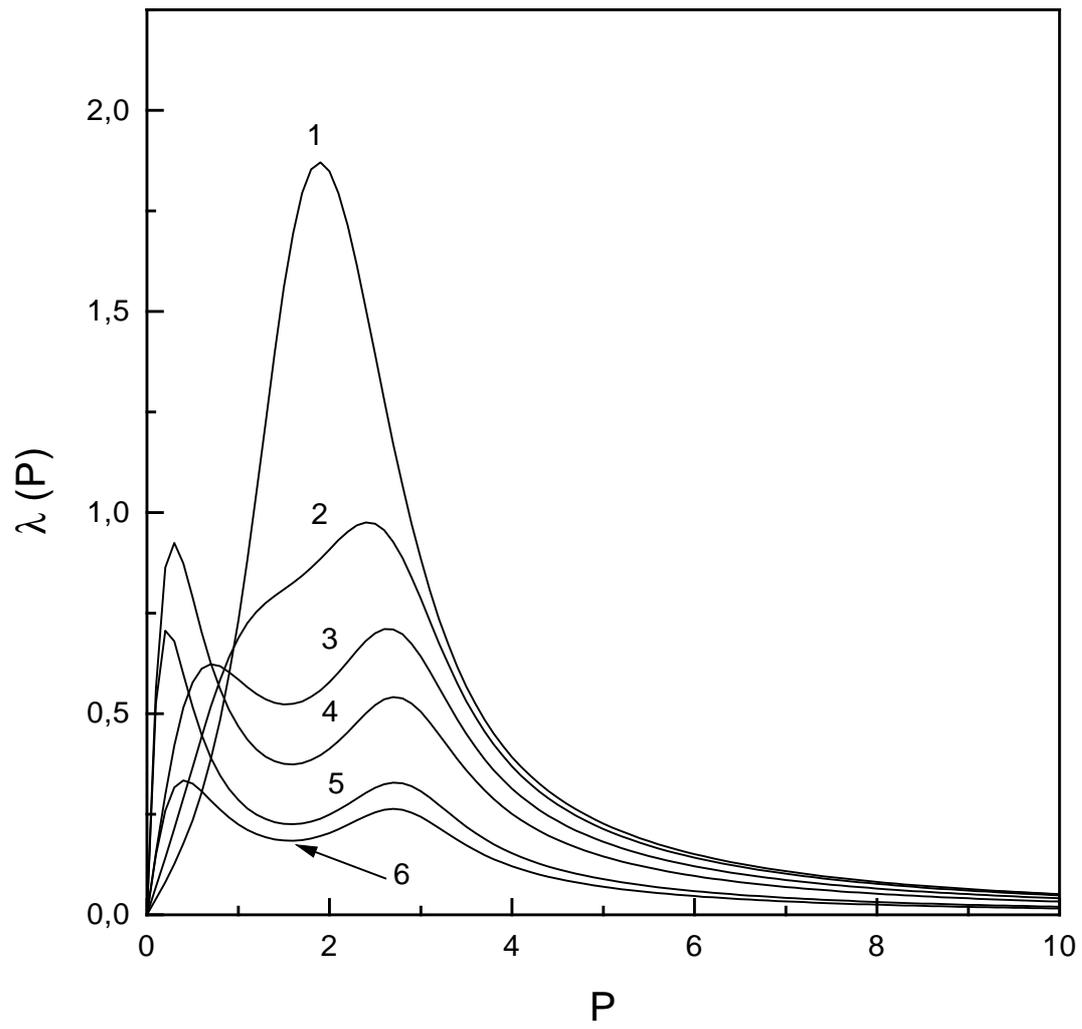
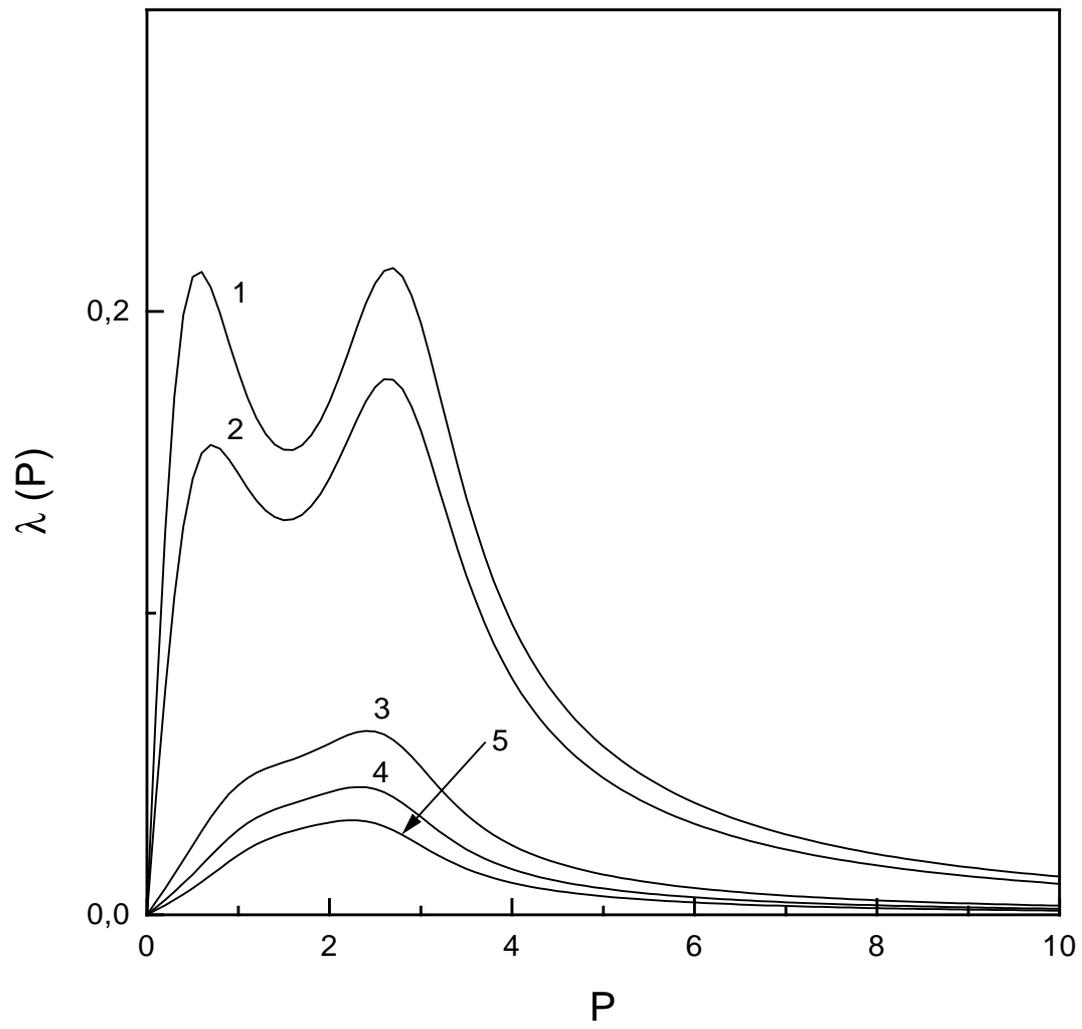
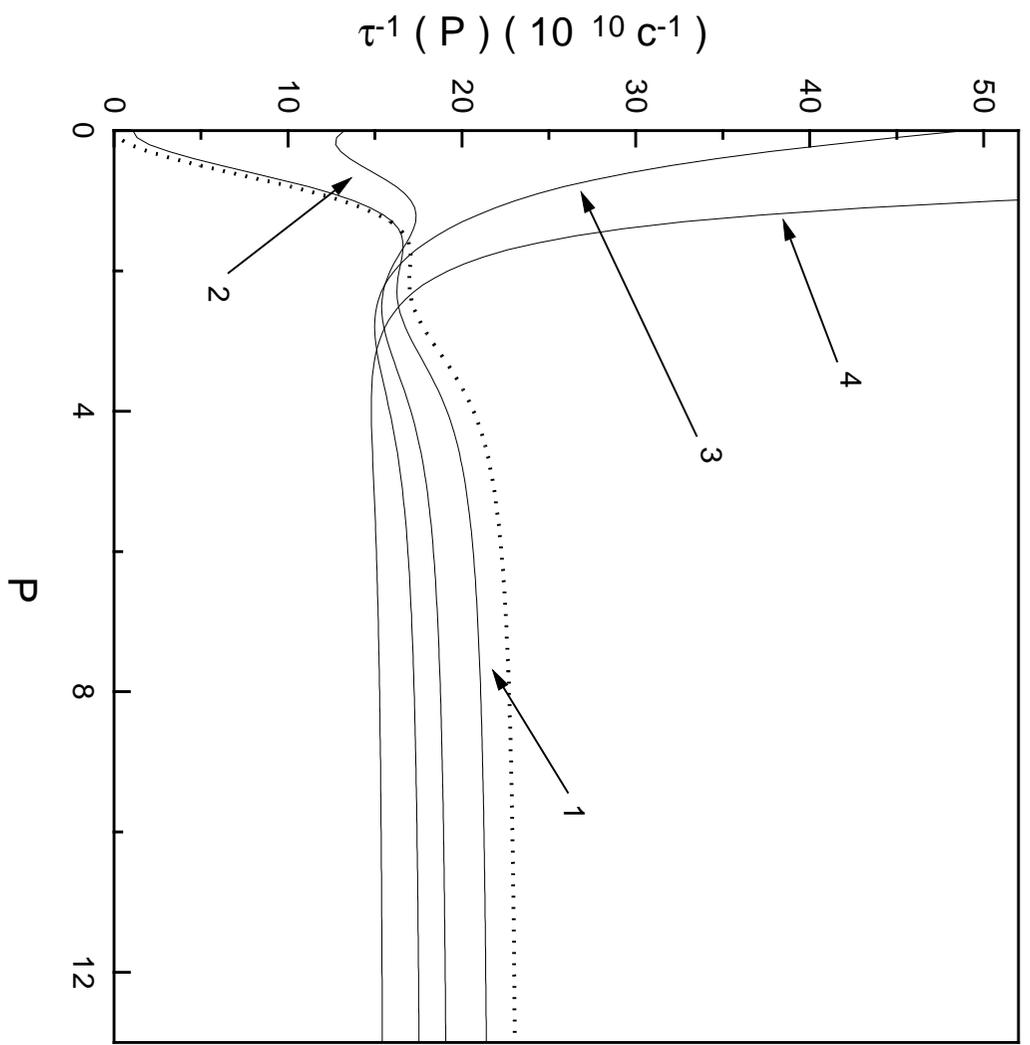
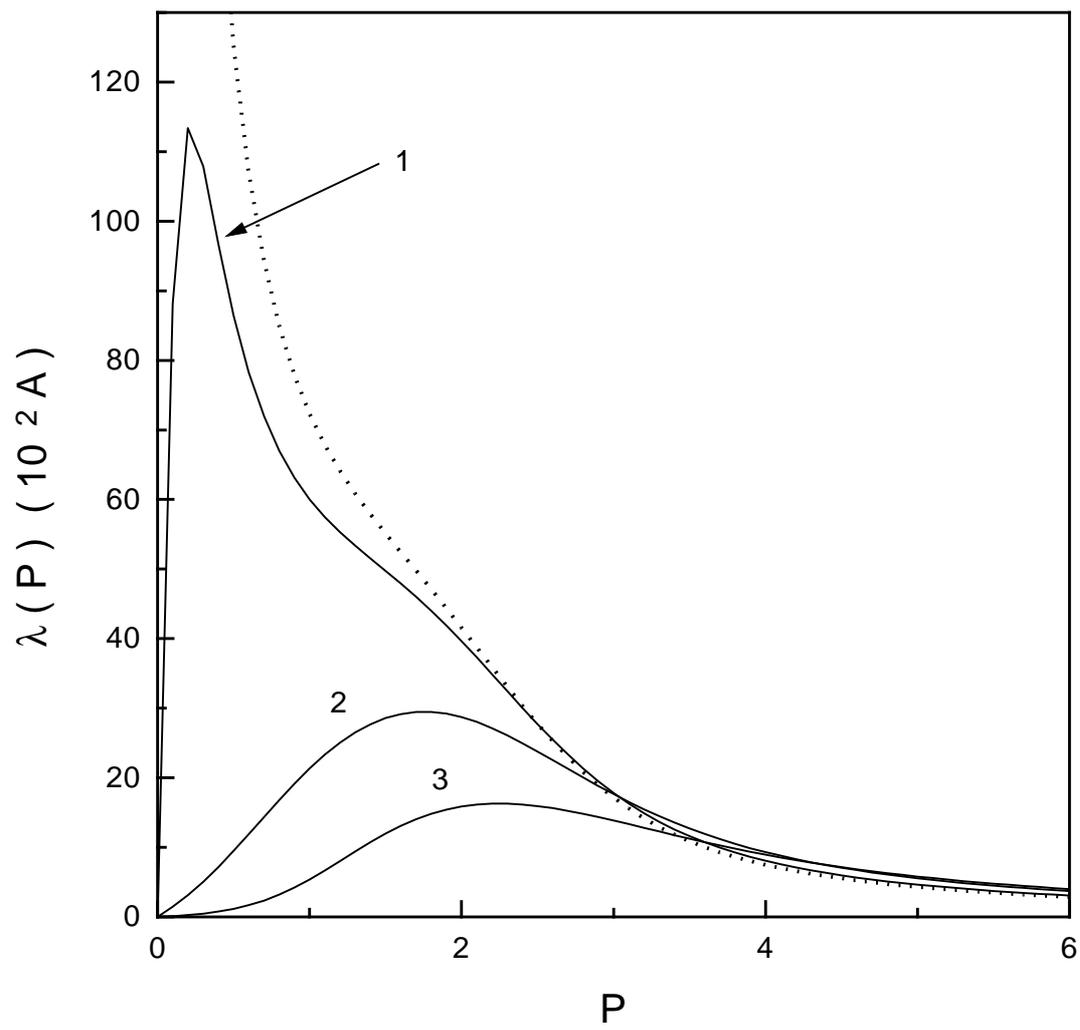
\end{document}